# Home Energy Management Systems in Future Smart Grids

I. Khan[1], A. Mahmood[1], N. Javaid[1], S. Razzaq[2], R. D. Khan[3], M. Ilahi[1]

[1]COMSATS Institute of Information Technology, Islamabad, Pakistan.
[2]COMSATS Institute of Information Technology, Abbottabad, Pakistan.
[3]COMSATS Institute of Information Technology, Wah Cant, Pakistan.

**ABSTRACT**
We present a detailed review of various Home Energy Management Schemes (HEM,s). HEM,s will increase savings, reduce peak demand and Pto Average Ratio (PAR). Among various applications of smart grid technologies, home energy management is probably the most important one to be addressed. Various steps have been taken by utilities for efficient energy consumption.New pricing schemes like Time of Use (ToU), Real Time Pricing (RTP), Critical Peak Pricing (CPP), Inclining Block Rates (IBR) etc have been been devised for future smart grids.Home appliances and/or distributed energy resources coordination (Local Generation) along with different pricing schemes leads towards efficient energy consumption. This paper addresses various communication and optimization based residential energy management schemes and different communication and networking technologies involved in these schemes.

## 1. INTRODUCTION

Smart grid is the integration of advanced information, communication and networking technologies in traditional electric grid to make it smarter and faster in making decisions.This integration will bring more automation, reliability of electrical services, safety of electrical equipments and hence an increase in consumer comfort level. With the advent of smart grid several emerging techniques and technologies have been proposed in past decade by researchers across the globe. Smart meters, bidirectional communication, advanced metering infrastructure (AMI), home automation and home area networks (HAN,s) are the techniques and technologies addressed by various researchers [1]. Traditional electric power grid has been serving humanity for the last one century. Population has increased, the traditional grid has worn out thus addition of more and more electric equipments bring instability to the traditional electric power grid [2].

Smart grid has applications in generation, transmission, distribution, and consumption of electrical energy. Smart grid technology enables distributed power generation, where power can be generated locally, use the required energy and sale extra power back to utility. In [3] a power quality monitoring strategy has been enabled by using sensor networks in smart grid (transmission & distribution application). Smart grid can also enhance the electricity usage efficiency. Consuming electrical energy efficiently has proved to be beneficial both economically and socially. By employing home energy management systems, a consumer can reduce his energy bill, reduce peak demand and hence contributes less towards environmental pollution by reducing emission of Green House Gases (GHG).

The demand curve in traditional power grid and flat pricing rates scenario shows that load demand is very high during peak periods as compared to off-peak periods.The result is that utility companies bring their peaker plants online which results in higher generation costs and emission of GHG,s. The originally inelastic demand curve needs to be changed to reduce energy cost and peak load demand. Home energy management (HEM) systems in smart grid enable Demand Side Management (DSM) and Demand Response (DR) programs. DSM is more related to planning, implementing and evaluating techniques and policies which are designed to modify the electricity consumption of consumers. Whereas DR programs are used to manage and alter energy consumption based on supply. Continuous efforts from research community are underway to design new protocols, standards and optimization methods to coordinate Distribution Energy Generation (DER) and residential appliances efficiently in order to reduce peak demand and energy consumption charges of consumers.

This paper addresses two types of home energy management schemes i.e one is communication based and the other is optimization based. The home energy management schemes are combined with different pricing schemes in order make the scheme more efficient.

2. **Energy Management**

Energy management is a broader term, which applies differently in different scenarios, but we are concerned about the one which is related with energy saving in homes, public sector/ government organizations or business. In this scenario the process of monitoring, controlling and conserving energy in an organization/ building may be termed as energy management [2]. In smart grid where the consumers can generate local energy from several distributive generation units and where there is a plenty of space for different pricing schemes, the need for energy management programs has been pointed out by many researchers. Previous work shows that energy management programs with feed in increased savings when compared with without feed in. Fig.1 [1] tells the story.

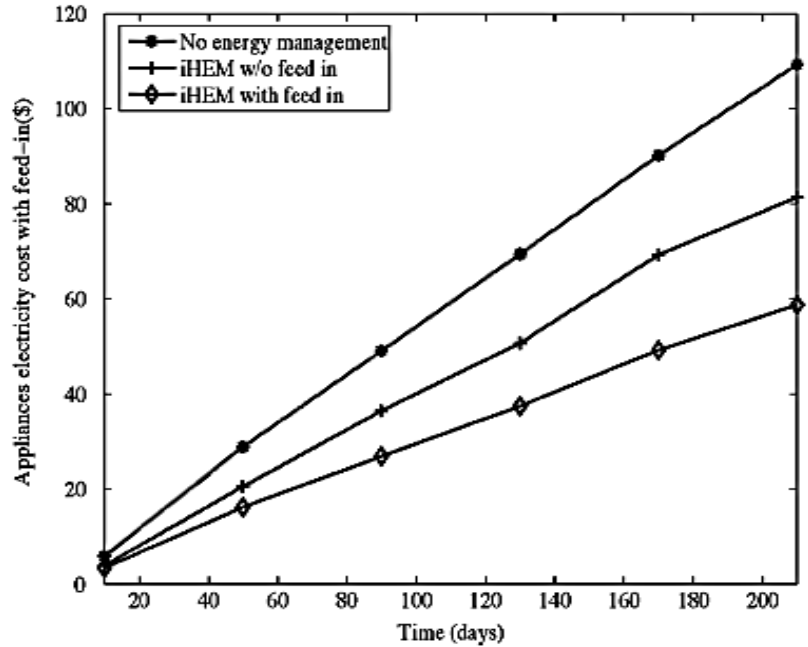
Figure 1: Appliance electricity costs

Demand side management can contribute in reduced emissions, reliable supply of power and lowering the energy cost. Current grid has demand side management programs for consumers like commercial buildings and industrial plants; however it does not have any such scheme for domestic consumers due to the reasons of lack of effective communication, efficient automation tools and sensors. Secondly implementation costs of various demand response programs are higher when compared with its impact. However in smart grid, smart loads, low cost sensors, smart meters and the information and communication technology open a window for domestic energy management programs [1]. Previous literature employs different techniques for energy management in smart grid.

**2.1 OREM**

Optimization based residential energy management (OREM) has been proposed by authors in [1]. The objective function given in (4) manages the energy consumption by scheduling home appliances in suitable timeslots. Scheduling an appliance in an appropriate timeslot may bring a non acceptable amount of delay to the appliance cycle and an exploded load in the low price timeslots. To tackle this problem an upper bound delay $D_{max}$ is specified for each appliance which is equal to the length of two timeslots. Methamatically

$$D_{max} \leq 2D_i \quad (1)$$

Where $D_i$ is the appliance operation cycle.

### 2.2 iHEM

In [1, 4] Wireless Sensor Networks (WSN,s) have been used for an in-home energy management (iHEM) application. The schemes are based on communication between smart appliances, a central Energy Management Unit (EMU), smart meter and storage unit. The algorithms applied have been designed for ToU pricing scheme, where electricity charges rates are different for different periods of time. iHEM application manages the home energy by shifting the load form peak to off-peak periods i.e. reduces peak load demand. When a consumer turns on an appliance a data packet is sent to the EMU. EMU then communicates with smart meter and local generation units to know about the price information from utility and the available local energy respectively. Based on these information, EMU schedules the starting time of the appliance. Waiting time of the appliance is calculated as the difference between the suggested time by EMU and request start time. The consumer requests have been modeled as a Poisson process. The simulation results show that the percentage contribution of the appliances to the total load during peak hours is reduced fig.1.

### 2.3 Decision Support Tool (DsT)

Decision Support Tool (DsT) has the primary aim to help users in making intelligent decisions during their appliance operation. Advantages of energy management program may be increased if besides appliance coordination there is distributed energy resources (DER) coordination too in parallel. In [5] the concept of DER coordination has been evaluated. The work has used an enhanced PSO solver i.e. CPSO-R, to quantify value added by the DER coordination. Coordination value has been calculated first for the case when each DER is scheduled independently and then for the case when the DER cooperates with each other.

### 2.4 Optimal and Automatic Residential Energy Consumption Scheduler

The optimization based residential load control scheme discussed in [6] is based on simple linear programming computations. The scheme is proposed for real time pricing which needs a price predictor. The combination of price predictor and energy consumption scheduling (ECS) design significantly lowers the peak to average ratio (PAR) in load demand for different load scenarios.

### 2.5 Optimum Load Management (OLM) Strategy

In [7] an optimization based residential load management strategy has been proposed. The optimization problem needs several interests forecasting and activity scheduling by users to form an objective function. Various interests are local power production i.e. from solar, wind etc, load, and electricity prices for next day. Following objective function is produced as a result [7].

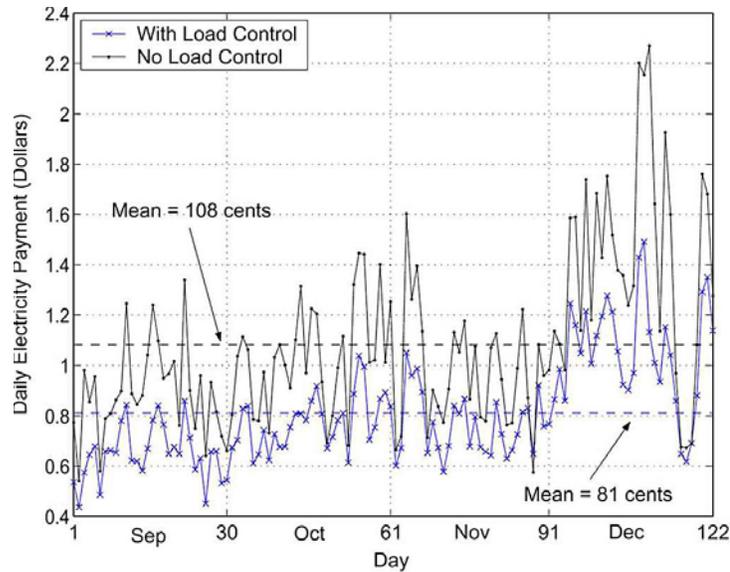

Figure 2: Trends of daily PAR for a typical residential load based based on DAP adopted by IPC from 1 September to 31 December 2009

### 3. Monetary Cost Minimization

Different pricing schemes may be employed for energy billing purposes by utilities like time of use (ToU) pricing scheme, real time pricing (RTP), day ahead pricing (DAP), critical peak pricing (CPP) etc. In ToU scheme, consumer is charged more during peak periods, less during mid-peak and the least during off-peak.Base plants provide power to base loads. During peak periods as the consumer demand goes higher the utilities switch on their peaker plants to maintain load-supply balance, which generate power at a comparatively high cost and with production of high amount of green house gases (GHG) [1]. The ultimate result is increase in price of electricity and global environmental problems. Smart grid gives opportunity to the consumers to generate local power from distributed energy resources(DER) e.g. solar, wind mill, use it locally and sell back the extra power to utility.Monetary cost minimization can be acheived by applying demand side management schemes. The energy management algorithms can shift load from peak periods to off-peak and hence reduce the cost and emission of GHG. Previous work shows that the objective can also be achieved by scheduling the distributed energy resources i.e. scheduling the DER by optimizing an objective function.In subsections below few of energy management schemes are presented.

#### 3.1 iHEM Application

iHEM, an energy management scheme is presented for domestic energy management in [1]. The scheme uses smart appliances, a central energy management unit (EMU) and wireless sensor home area networks (WSHAN) for communication purposes among appliances, EMU and smart meters. iHEM uses Zigbee protocol for the implementation of wireless sensor network, organized in cluster-tree topology. The application is based on appliance coordination system (ACS). Unlike the OREM the consumer demands are processed in near real time in iHEM.

The consumer may turn on any appliance at any moment on the clock irrespective of the peak hours concern and iHEM suggest a convenient start time to the consumer. On switching the appliance on, a START-REQ packet is sent by the appliance to the EMU. Upon receiving the START-REQ packet, EMU communicates with the storage system to inquire about the available stored energy by sending AVAIL-REQ packet.EMU also communicates with smart meter to know about the updated prices. The storage unit sends an AVAIL-REP packet in reply, containing the information about the amount of stored energy. When EMU receives the AVAIL-REP packet, it schedules a convenient start time for the appliance according to the iHEM algorithm and notifies it to consumer by sending a START-REP packet. The consumer, at this stage may be willing to negotiate with EMU, through the NOTIFICATION packet.Message flow is shown below in fig.3. [1] for iHEM application.

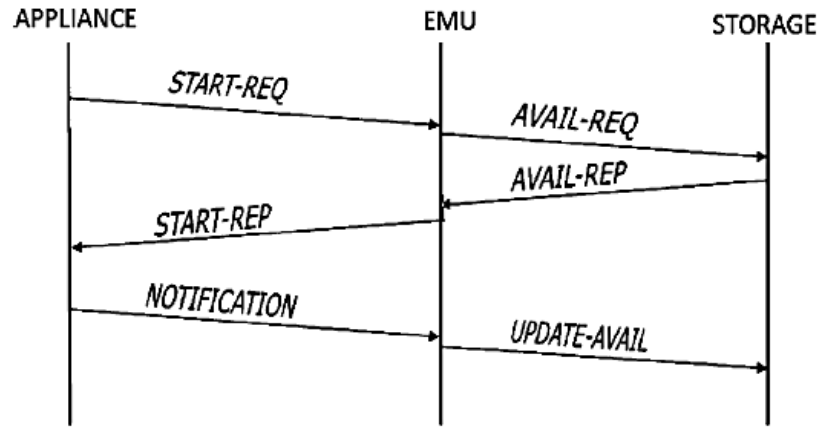

Figure 3: Message flow in iHEM application

30% of the load takes place during peak hours in the absence of energy management programs. By employing iHEM, peak load can be reduced upto 5% [1]. The simulation results shows that iHEM reduces carbon emission and energy consumption costs.

### 3.2 ACORD

In [8] ACORD scheme has been proposed to benefit from ToU pricing and decrease energy cost. Aim of ACORD scheme is to shift the consumer load to off-peak periods. In-home WSN's are used for delivery of consumer requests to EMU. The work shows that the rate of consumer requests has a sizeable effect on energy cost reduction. Energy consumption lowers significantly with an increase in request rates from consumer side. This scheme only considers the scheduling of home appliances. Simulation results have been shown for energy cost reduction with ACORD.

### 3.3 ACORD-FI

ACORD-FI [9] is another energy management scheme for energy-aware smart homes. In ACORD-FI both the home appliances and distributed energy resources are scheduled with the purpose of reducing the energy bill and green house gases [9]. ACORD-FI schedules consumer requests considering peak hours, local energy generated and other conflicting requests. ACORD-FI uses WSN,s for communication between EMU and appliances and smart meters.

### 3.4 Decision Support Tool (DsT)

In [5] the authors have devised a DsT for smart homes. The DsT based scheme coordinates only the distributed energy resources (DER). As acase study, a space heater, PHEV, pool pump, PV system and a water heater are scheduled based on various ToU prices by applying the coevolutionary particle swarm optimization (CPSO) technique.A typical smart home case study is shown in the figure.4 [5].

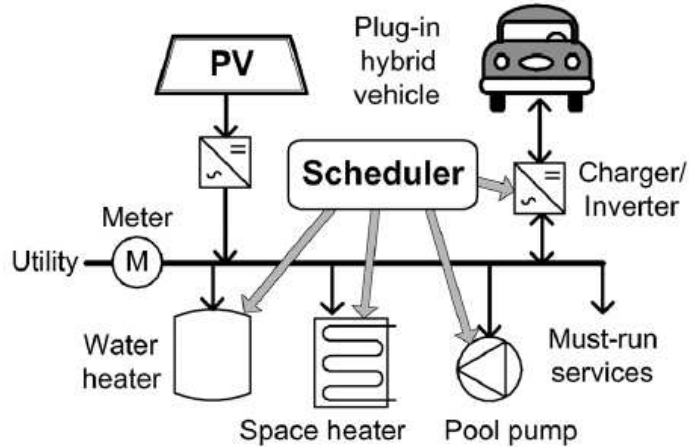

Figure 4: A DsT smart home case study

DsT is composed of DER scheduling algorithm and an energy service model. The net benefit of to the consumer is maximized by scheduling the controllable DER according to the scheduling algorithm. And the consumer energy consumer bill may be reduced by 16-25% [5].

### 3.5 Residential Load Control (RLC)

In [6] an optimal and automatic residential energy consumption scheduler has been proposed for a scenario where real time pricing (RTP) is combined with inclining block rates (IBR). Load control programs for real time pricing needs a price predictor hence the scheduler is combined with a price predictor shown in fig.5. The simulation results show that this combination reduces the consumer's payments by 10-25%.

### 3.6 Optimum Load Management (OLM)

Using the communication infrastructure of the future smart, this paper proposed an optimal load management strategy for real time pricing. By adopting this strategy, consumers can bring a balance between their energy bills and economical situations. The primary purpose of this strategy is to reduce the energy consumption cost of the consumers. Simulation results show that OLM can reduce energy bill by 8-22% [7].

### 4. Communication and Networking Technologies for Energy Management Applications in Smart Grid

Integration of communication and information technologies (ICT) in electrical power grid will transform the current grid into smart grid [10]. Efficient communication, reliable networking and better control will bring safety, reliability, improved efficiency and sustainability of electrical services. ICT integration may enhance capabilities of various applications of smart grid and home energy management (HEM) is one of them.

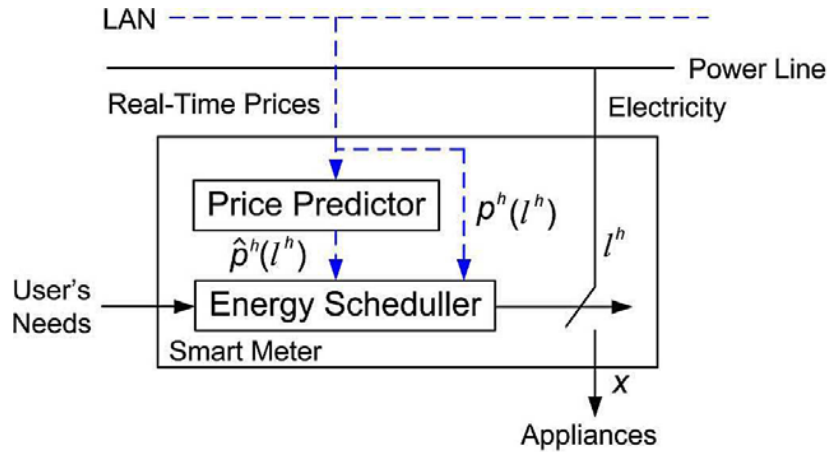

Figure 5: A design for RLC scheme

Previous literature shows how the communication, networking and control capabilities of smart grid can be used in generation, transmission, distribution and consumption of electrical power. At consumption level, HEM employs home area networks (HAN,s) for inter networking of smart loads, energy management units and smart meters. Various communication technologies both wired and wireless have been proposed for bidirectional communication in smart grid. A brief summary about the communication and networking technologies suggested for HAN in smart grid has been shown in fig.6.

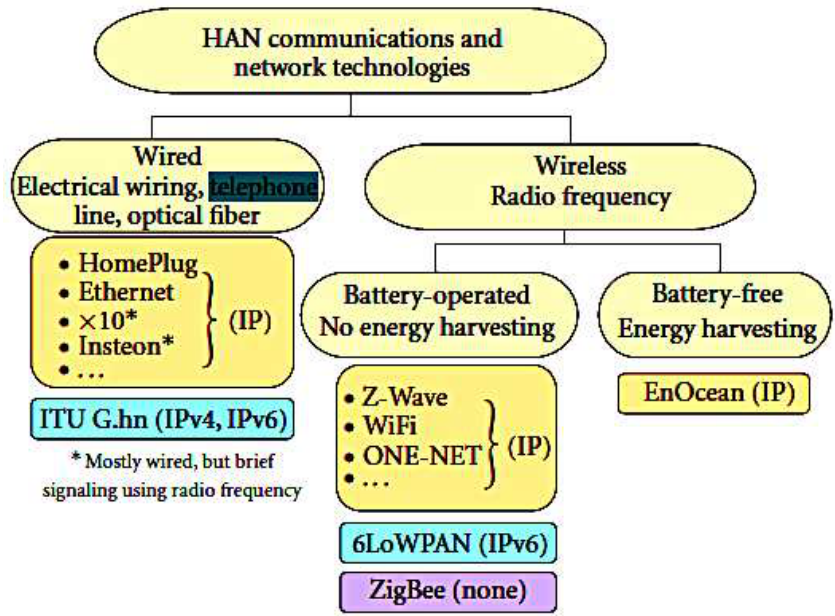

Figure 6: Communication and Networking Possibilities in HAN

HAN,s comprise of two types of system. One is the command based system where a very short acquisition time is required for data exchange and hence a low data rate and bandwidth is required. HEM application is associated with command based system. The other one is called link-based system (communication and entertainment) which needs a reliable communication (point-to -point) and hence the requirements for data rate and bandwidth goes higher e.g. multimedia streaming, voice. A multidimensional graph is presented in fig.7 for the comparison of the two systems described above. It is shown that HEM application costs the least and different communication technologies have been placed at its proper place on the basis of data rates requirements.

Wireless sensor home area networks (WSHAN,s) have been deployed in various smart home projects for monitoring and supervision purposes. In [1] an appliance coordination based home energy management application uses the in-home wireless sensor network implementing the Zigbee protocol for energy management purposes.

Zigbee is an IEEE 802.15.4 based wireless technology with a typical range of 10-75 m, low cost, and with a low power usage. Two types of devices are allowed by Zigbee. i.e. Full function devices (FFD,s) and reduced function devices (RFD,s).

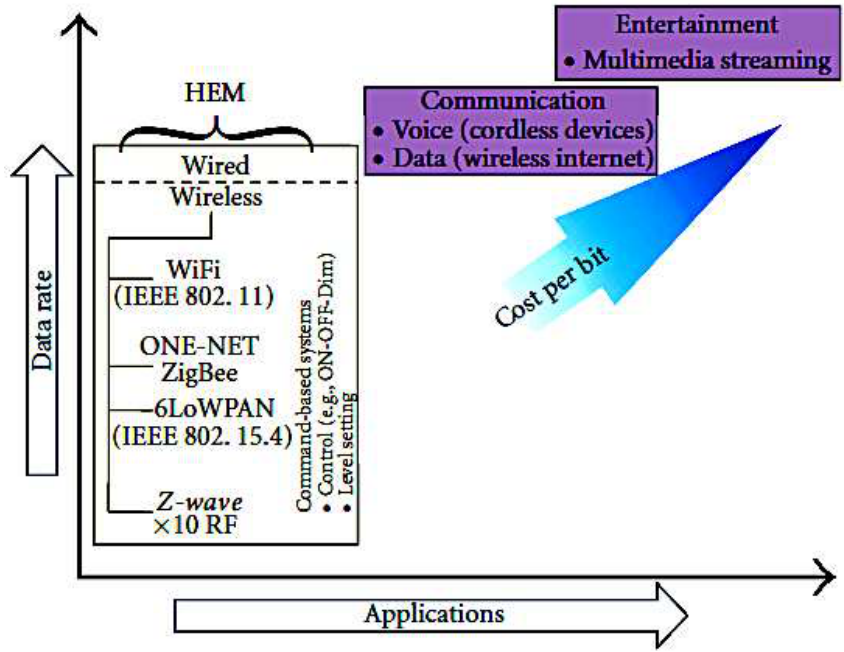

Figure 7: Comparison of Command based system and Link based system

FFD,s are interconnected in a mesh topology which means it can communicate with its peers. On the other hand RFD,s can be edge nodes in a star topology where they cannot communicate with its peers. In iHEM model home the wireless sensor home area network is organized in a cluster-tree topology as shown in fig.8.

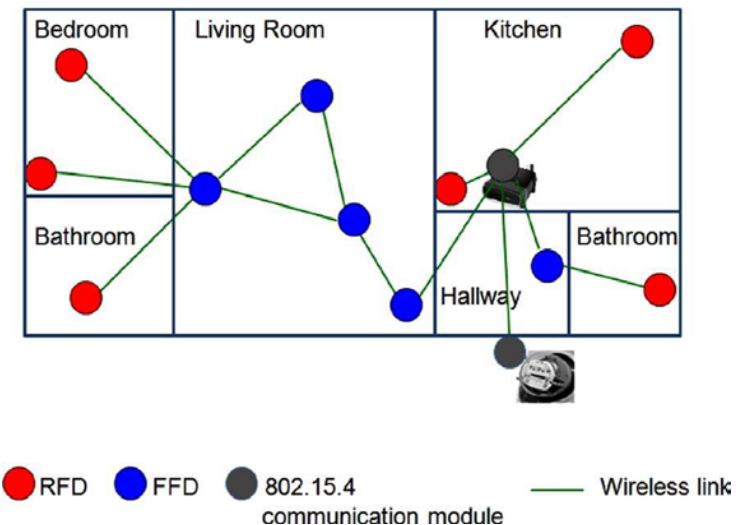

Figure 8: WSHAN Topology in iHEM

5. Conclusion

On a conclusion note, this paper has revisited the need for domestic energy management for efficient consumption of electricity in smart grid. Consuming electrical energy efficiently results in reducing peak load, lowering electricity bills and minimizing the emission of green house gases (GHG). In smart grid where there is bidirectional communication and better home automation, effective home energy management system can be designed. This paper has discussed several home energy management schemes where different pricing schemes have been applied to get economical and social advantages. Both communication-based and optimization-based home energy management techniques have been evaluated. We have also discussed some communication and networking technologies for future smart grid that can play a key role in smart energy usage systems in future smart grids. We are of the hope that this work will channelize the efforts towards a more efficient, user friendly home energy management system for future smart grids.